# A UNIVERSAL SCALING LAW FOR NANOINDENTATION, BUT NOT ONLY

Nicola M. Pugno
Department of Structural Engineering, Politecnico di Torino,
Corso Duca degli Abruzzi 24, Torino, 10129, Italy

In this letter we derive a universal law for nanoindentation, considering different sizes and shapes of the indenter. The law matches as limit cases all the well-known hardness scaling laws proposed in the literature. But our finding can also explain their deviations experimentally observed at the nanoscale. An even more general scaling law is then formulated, also in the fast and slow dynamics; it is based only on the surface over volume ratio of the domain in which the energy flux occurs: thus, its application in different fields, also for chaotic and complex (e.g., biological) systems, is demonstrated.

Hardness is defined as the load divided by the projected area of the indentation, thus it is the mean pressure that a material will support under load. This parameter is only "nominally" a constant and is experimentally dependent on penetration depth, size and shape of the indenter. A variation in hardness versus penetration depth is usually defined, and perhaps not properly, as indentation size-effect, whereas we should refer to the variation of hardness by varying the size or shape of the indenter as true size- or shape-effects respectively. Much of the early work on indentation was reviewed by Mott (1956). Ashby (1970) proposed that geometrically necessary dislocations (Nye, 1953) would lead to an increasing in hardness measured by a flat punch. The problem to a conical indenter has been investigated by Nix and Gao (1998), showing a good agreement with microindentation experiments. However, recent results that cover a greater range of depths show only partial (Poole et al., 1996; Swadener et al., 2002) or no agreement (Lim and Chaudhri, 1999) with their model. Thus, such a model was extended by Swadener et al. (2002) in a very interesting way, to treat indenters of different sizes and shapes; the results were compared with microindentation experiments, but limitations for small depths of pyramidal indenters and sizes of spherical indenters were observed at the smaller size-scales. The aim of this letter is the developing of a new model capable of matching as limit cases the discussed indentation laws, simultaneously capturing the deviation observed at the nanoscale. Incidentally, an extension of the Taylor's law (1938) is formulated. Moreover, a very general scaling law is provided, that seems to promise interesting applications in different fields, such as engineering, physics, biology, medicine, economy, to cite a few, also for complex and chaotic systems and also in the fast and slow dynamics.

Consider an indenter with a given geometry $h = h(r, \vartheta)$, with $r$ and $\vartheta$ polar coordinates. The Nix and Gao (1998) and Swadener et al. (2002) models assume that plastic deformation of the surface is accompanied by the generation of geometrically necessary dislocation loops, in our treatment of length $l(h)$, below the surface; the deformation volume is assumed to be an hemispherical zone below the contact area $A$ with radius $a = \sqrt{A/\pi}$ or volume $V = 2\pi/3 (A/\pi)^{3/2}$ (Figure 1). Thus, the total length $L$ of the geometrically necessary dislocation loops can be evaluated by integrating the number of steps on the indented surface (see Figure 1):

$$L = \frac{1}{b}\int_0^h l(h)\,dh = \frac{S}{b} \tag{1}$$

where $S$ is the lateral surface area of the indented zone, or of the indenter itself for monotonic positive defined variation $h$ vs. $r$ (Figure 1) and $b$ is the (modulus of the) Burger's vector. Note the generality of the result in eq. (1), that do not need the specification of the form of $h$, as required by the previous models. Thus, the average geometrically necessary dislocation density is:

$$\rho_G = \frac{L}{V} = \frac{S}{bV} \tag{2}$$

Arsenlis and Parks (1999) have shown that the actual number of dislocations that must be generated to accommodate plastic deformation is greater than the number of geometrically necessary dislocations by the so-called Nye factor $\bar{r}$ (~2, Swadener et al., 2002), thus, the total dislocation density is:

$$\rho_T = \bar{r}\rho_G + \rho_S \tag{3}$$

where $\rho_S$ is the statistically stored dislocation density. Previous works (Nix and Gao, 1998; Swadener et al., 2002) assume $\rho_T$ related to the shear strength $\tau_p$ by the Taylor (1938) hardening model, i.e., $\tau_p = \alpha\mu b\sqrt{\rho_T}$, where $\mu$ is the shear modulus and the geometrical constant $\alpha$ is usually in the range 0.3-0.6 for FCC metals (Wiedersich, 1964). However we note here that this law cannot be considered reasonable at the nanoscale, where $\tau_p$ must approach the nanoscale material strength $\tau_p^{(nano)}$, that, at the truly atomic scale, must be coincident with the theoretical material strength. Thus, the natural generalization of the Taylor's law is straightforward:

$$\tau_p = \frac{\alpha\mu b}{\sqrt{\rho_T^{-1} + \beta b^2}}, \quad \beta = \left(\frac{\alpha\mu}{\tau_p^{(nano)}}\right)^2 \tag{4}$$

The correct interpretation of eq. (4) is the definition of the real (thus finite) total dislocation density $\rho_T'$ as $\rho_T'^{-1} = \rho_T^{-1} + \beta b^2$, that, as a consequence of the quantized nature of matter, must have an upper bound of the order of $b^{-2}$, as for a pure single dislocation. This is also reflected in the expression of $\beta$, noting that $\alpha\mu$ is of the same order of magnitude of the theoretical material strength. According to this interpretation the Taylor's law survives, but considering the substitution $\rho_T \to \rho_T'$ for accounting the discrete nature of the energy dissipation. Note the analogy with quantized fracture mechanics (Pugno and Ruoff, 2004; Pugno 2006a), that quantizing the crack advancement, as must be at the nanoscale, predicts a finite theoretical material strength, in contrast to the result of the continuum-based linear elastic fracture mechanics (Griffith, 1920).

Assuming that the flow stress is related to the shear strength by the von Mises' rule, i.e., $\sigma_p = \sqrt{3}\tau_p$, and the hardness to flow stress by the Tabor (1951) factor of 3 (Nix and Gao, 1998; Swadener et al., 2002), the model gives the following hardness prediction:

$$\frac{H(S/V)}{H_{nano}} = \left(\frac{\delta^2-1}{\ell S/V+1}+1\right)^{-1/2}, \quad \frac{H(S/V)}{H_{macro}} = \left(\frac{\delta^2-1}{\delta^2 V/(\ell S)+1}+1\right)^{1/2}, \quad \delta = \frac{H_{nano}}{H_{macro}} \qquad (5)$$

with $H_{nano} \equiv H(\ell S/V \to \infty) = 3\sqrt{3/\beta}\alpha\mu$ (at the atomic scale expected to be coincident with the theoretical material hardness), $H_{macro} \equiv H(\ell S/V \to 0) = \dfrac{3\sqrt{3}\alpha\mu b}{\sqrt{\rho_S^{-1}+\beta b^2}}$ and $\ell = \dfrac{\bar{r}}{\rho_s b}$, i.e., a characteristic length governing the transition from the nano- to the macro-scale. The two equivalent expressions in eq. (5) correspond to a *bottom-up* or a *top-down* view, even if the bottom-up law is perhaps more physical, starting from the intrinsic material property $H_{nano}$. Eq. (5) is the mentioned universal law for nanoindentation, that provides the hardness as a function only of the ratio between the surface throughout the energy flux propagates and the volume where the energy is dissipated; or, let simply say, as a function of the surface over volume ratio of the domain in which the energy dissipation occurs. Such a law can be applied in a very simple way to treat any interesting indenter geometry. However, to make a comparison, let us focus on the axially symmetric profiles, i.e., $h = h(r)$, investigated by Swadener et al. (2002).

*Conical indenter.* Considering a conical indenter with corner angle $\varphi$, $h(r) = \tan((\pi-\varphi)/2)r$; geometrically we found $S/V = \dfrac{3\tan^2((\pi-\varphi)/2)}{2h}$, that introduced into eq. (5) gives $H_{cone}(h,\varphi) = H_{macro}\sqrt{1+\dfrac{(1-\delta^{-2})h^*(\varphi)}{h+\delta^{-2}h^*(\varphi)}}$, with $h^*(\varphi) = 3/2\ell\tan^2((\pi-\varphi)/2)$. For $h/h^* \to \infty$ or $\varphi \to \pi$ $H_{cone} \to H_{macro}$, whereas for $h/h^* \to 0$ or $\varphi \to 0$, $H_{cone} \to H_{nano}$; only for the case of $\delta \to \infty$, $H_{cone} = H_{macro}\sqrt{1+h^*/h}$ as derived by Nix and Gao (1998). Note that such a scaling law was previously proposed by Carpinteri (1982) for material strength (with $h$ structural size). But the comparison here is not very significant, since the indentation size-effect is not a true size-effect.

*Parabolic (spherical) indenter.* Consider the case of a parabolic indenter with radius at tip $R$, i.e., $h = r^2/(2R)$, that for not too large indentation depth corresponds also to the case of a spherical indenter; geometrically we found $S/V = 1/R$, that introduced into eq. (5) gives $H_{parabola}(R) = H_{macro}\sqrt{1+\dfrac{(1-\delta^{-2})R^*}{R+\delta^{-2}R^*}}$, with $R^* = \ell$; thus, the hardness should not be a function of the indentation depth $h$. For $R/R^* \to \infty$, $H_{parabola} = H_{macro}$, whereas for $R/R^* \to 0$, $H_{parabola} = H_{nano}$; only for the case of $\delta \to \infty$, $H_{parabola} = H_{macro}\sqrt{1+R^*/R}$, as derived by Swadener et al. (2002). This law describes a true size-effect and agrees with the Carpinteri's law (1982).

*Flat indenter.* Consider the case of a flat indenter of radius $a$, i.e., $h = \delta(r-a)$; geometrically we found $S/V = \dfrac{2\pi ah+\pi a^2}{2/3\pi a^3}$, that introduced into eq. (5) gives the expression of $H_{flat}(a,h)$. For $a/\ell \to \infty$, $H_{flat} \to H_{macro}$, whereas for $a/\ell \to 0$, $H_{flat} \to H_{nano}$; interestingly for $h/\ell \to \infty$ $H_{flat} \to H_{nano}$, showing an inverse indentation size-effect, in agreement with the discussion by Swadener et al. (2002) and with the intuition (the contact area does not change when the penetration load or depth increases), see Ashby (1970). This suggests a new intriguing methodology to derive the nanoscale hardness of materials by a macroscopic experiment, using large flat punches (but the finite curvature at the corners is expected to affect the results). On the other hand, for $h/\ell \to 0$ (or $h \propto a$) and $\delta \to \infty$, $H_{flat} = H_{macro}\sqrt{1 + a^*/a}$, with $a^* = 3/2\,\ell$. This case was only discussed by Swadener et al. (2002), due to the complexity in their formalism to treat such a cuspidal geometry. Note that the last size-effect law is again coincident with that of Carpinteri (1982).

The generality of the formulation suggests us that the law of eq. (5) can have large applications. One example is in the design of nanosyringes, e.g., composed by a single walled carbon nanotube, for which the force is predicted to be $F_{nano\atop syringe} \approx \pi D s H_{nano}$, where $D$ is the nanotube diameter and $s$ is the wall thickness; accordingly, the maximum length to avoid elastic instability of the nanosyringe will be $l_{max} \approx \pi \sqrt{EI/F_{nano\atop syringe}}$, where $E$ is the Young's modulus and $I$ is the moment of inertia of the nanotube. For example, considering the following reasonable parameters $D$=10nm, $s$=0.34nm, $H_{nano} = 10\text{GPa}$, $E = 1\text{TPa}$, would result in $F_{nano\atop syringe} \approx 107\text{nN}$ and $l_{max} \approx 111\text{nm}$. An additional example is on the design of bullet-proof rackets: the work needed to penetrate into the material for a length $l$ is $W = \int_0^l A(h)H(h)\mathrm{d}h = l\langle AH \rangle$, where $A$ is the cross-section bullet projected area; since $W$ must be equal to the bullet kinetic energy $K$, its minimum thickness $l$ is predicted to be $l_{min} = \dfrac{K}{\langle AH \rangle}$. For example, considering a bullet mass of 100g, velocity of 1Km/s, $A$=1cm$^2$ and $H$=10GPa, would correspond to $l_{min} \approx 5\text{cm}$.

Note that defining the impact strength as the energy spent over the removed volume, the last example shows that such a parameter is for plastic materials of the order of the hardness, but note: at the investigated size-scale (and $H_{nano} >> H_{macro}$). A similar result is found for brittle materials at the macroscale, where the impact strength is found to be of the order of the material macro-strength (Pugno, 2006b): thus, our argument suggests its validity at all the size-scales by considering the related material strength. Summarizing for plastic material the impact strength is of the order of the hardness whereas for brittle material of the material strength, but at the considered size-scale. This finding could have interesting applications also in impact or explosion and tribological studies, from the nano- to the macro-scale. Other examples of applications could be done; but let us return to the nanoindentation problem.

Swadener et al. (2002) compared their model with experiments in annealed iridium, using spherical indenters of different radii ($R \approx 14, 69, 122, 318, 1600\,\mu\text{m}$). Data were analyzed using the Oliver and Pharr (1992) method. Swadener et al. (2002) treated the spherical indentation with their law for parabolic indenter, since those experiments were performed at a small value of penetration ($a/R \approx 0.05$). A deviation for $R < 80\,\mu\text{m}$, i.e., a lower hardness with respect to their prediction, was experimentally observed. This deviation is in agreement with the prediction of eq. (5). In particular, Swadener et al. (2002) considered two plausible sets of parameters for describing their experiments (see their motivations for details): (a) $H_{macro} \approx 0.9\text{GPa}$, $R^* \approx 250\,\mu\text{m}$ or (b) $H_{macro} \approx 0.6\text{GPa}$, $R^* \approx 750\,\mu\text{m}$. Introducing such values into our model with in addition (a) $H_{nano} \approx 3\text{GPa}$ or (b) $H_{nano} \approx 5\text{GPa}$ result in a closer agreement between theory and experiments, as shown in Figure 2 (a) and (b) respectively. Similar results with the same sets of parameters were observed for pyramidal indenter (Berkovich, treated as a conic; from its geometry $R^*/h^* \approx 5.2$, Swadener et al. 2002) on the same material, by varying the indentation depth. Annealed oxygen-free copper tested with spherical, Berckovich and Vickers (pyramidal, $R^*/h^* \approx 5.2$) indenters resulted in $H_{macro} \approx 0.1\text{GPa}$ and $R^* \approx 200\mu\text{m}$, but with the expected deviation at the smaller size-scales, again well-predicted by our model with $H_{nano} \approx 2\text{GPa}$. The same but cold-worked material was similarly investigated, giving $H_{macro} \approx 0.9\text{GPa}$ and $R^* \approx 3.6\mu\text{m}$, with a deviation again well described by the finite value $H_{nano} \approx 2\text{GPa}$.

Furthermore, we note that at the truly nanoscale we expect even higher values for $H_{nano}$. A tendency to very high values, observed by a reduction in the classical slope -1/2, has been observed in the indentation hardness of surface Si(111) films with an indentation depth as small as 1 nm (Bhushan and Koinkar, 1994). The hardness for an indentation depth of 2.5 nm is 16.6 GPa and drops to to 11.7 GPa at a depth of 7 nm, from which we deduce a slope of -0.34 (and thus $H_{nano} > 20\text{GPa}$).

However, our treatment is not restricted to nanoindentation. Infact, for a given nominally constant property *P* the generalization of eq. (5) is straightforward:

$$P(S/V) = P_{nano}\left(\frac{(P_{nano}/P_{macro})^\gamma - 1}{\ell\, S/V + 1} + 1\right)^{-1/\gamma} \tag{6a}$$

in which the parameter $\gamma$ has to be introduced, since a material property could also for example be $P = H^2$ (for which $\gamma = 1$, to match eq. (5)) or $P = H^{-1}$ ($\gamma = -2$), or others. For steady state energy flux *S* and *V* are per unit time. The asymptote at $P_{macro}$ is classical, and conceptually intrinsic in considering a nominally constant property, whereas that at $P_{nano}$ appears as a consequence of the existence of a nanoscale quantization. The law of eq. (6a) can be applied for predicting the scaling of a given property, starting from the surface over volume ratio of the domain in which the energy exchange or flux, not necessary a dissipation, occurs. For material strength $P = \sigma$, $\gamma = 2$ (as for *H*), and considering self-similar structures, i.e., $V/S \propto L$, as the characteristic structural size (but note that in general eq. (6a) describes also the shape-effects), we deduce $\sigma(L) = \sigma_{macro}\sqrt{1 + \frac{\ell^*}{L + \ell'}}$ with $\ell^* = (1 - \delta^{-2})\ell$, $\ell' = \delta^{-2}\ell$ and $\delta = \sigma_{nano}/\sigma_{macro}$ (smaller is stronger). This is a nano-scaling law, taking into account the quantization of the energy flux, and for $\delta \to \infty$ agrees with the Carpinteri's law (1982) (correspondence principle), that has already been demonstrated to agree with microtorsion (Fleck et al., 1994) and microbend (Stolken and Evans, 1998) experiments, see Gao et al. (1999a,b) and Huang et al. (2000). An infinity of other examples could be mentioned, e.g., scaling of fracture energy, friction coefficient, wear resistance, elastic modulus, and others, to cite a few in the material science.

The law of eq. (6a) can be applied also to complex and chaotic systems, where the multiscale energy flux arises in a fractal domain of positive dimension *D* (usually comprised between 2 and 3, i.e., between an Euclidean surface and volume); in this case $S \propto V^{D/3}$ (Carpinteri and Pugno, 2002), no matter if we are considering fractal fragments or dislocations, thus brittle or plastic materials (Carpinteri and Pugno, 2005). Accordingly, $S/V \propto L^{D-3}$ in eq. (6a) with *L* structural size. For example, for a hierarchical material (as for bone and dentine) we derive $D = 3\ln n/\ln(n/\varphi)$, where $n > 1$ and $0 < \varphi < 1$ are the number and volumetric fraction content of sub-inclusions in a main inclusion (the demonstration is left up to the reader). One practical example is given by the scaling of the energy density $P = \psi$ (nominally a material constant) during fragmentation of solids, for which $\gamma = 1$ (since $\psi \propto H^2$), in agreement at the intermediate size-scales with the mesoscopic scaling $\psi \propto S/V \propto L^{D-3}$ (Carpinteri and Pugno, 2002). Interestingly, such a law is also extensively applied for describing the scaling of the energy per unit mass spent by biological systems on growth (West et al., 2004; Delsanto et al., 2004; Carpinteri and Pugno, 2005): thus, eq. (6a) straightforwardly extends this biological scaling law, as well as the large number of allometric biological laws that can be derived from it (West et al., 2004). An infinity of other applications could be mentioned, as the scaling of the efficiency of nanoparticle-based rocket propellants, (fractal particle distribution), of absorption or corrosion and of other surface properties.

The last application is in the field of fast and slow dynamics (Delsanto et al., 2005). For a wave (on in general signal) propagation the size-scale $L$ is connected to the time-scale $t$ by $L \propto t$ (smaller is faster); in this context eq. (6a) is rewritten as ($S/V \propto 1/L \propto 1/t$):

$$P(t) = P_{fast}\left(\frac{\left(P_{fast}/P_{slow}\right)^\nu - 1}{\tau/t + 1} + 1\right)^{-1/\nu} \tag{6b}$$

where $\tau$ is a characteristic time.

An example for the material science community is on the variability of the dynamic strength $P = \sigma$ as a function of the time to failure $t$. The impact strength ($\sigma_{fast}$) is observed for the majority of the systems to be approximately twice the static strength ($\sigma_{slow}$); this seems to be related to the existence of an incubation time ($\tau$) for fracture nucleation, of the order of the time needed to generate a fracture quantum and thus again related to a quantization (Pugno, 2006a). It is evident that eq. (6b) with $\sigma_{fast} \approx 2\sigma_{slow}$ catches the described phenomenon (quantized dynamic fracture mechanics would suggest $\nu = 1/2$, see Pugno, 2006). The Young's modulus has a similar scaling as a consequence of visco-plastic activation, but an infinity of other applications are straightforward, as on the conditioning and frequency shift in the fast and slow dynamics (Delsanto et al., 2005) or earthquakes triggering, similar to an crack propagation incubation time, but arising at the megascale. For complex and chaotic fractal systems, the result previously reported ($S/V \propto L^{D-3}$) is formally traduced in eq. (6b) as $\tau \propto t^{D-2}$, where $D$ is here connected to the fractal dimension of the time distributions.

To formulate an even more general universal spatial-temporal scaling law $P(S/V, t)$ we have to combine eqs. (6a) and (6b); two different ways can be followed: considering $P_{nano}$ and $P_{macro}$ in eq. (6a) as a function of time and of $P_{nano \atop fast}$, $P_{nano \atop slow}$, $P_{macro \atop fast}$, $P_{macro \atop slow}$ according to eq. (6b) or, complementary, considering $P_{fast}$ and $P_{slow}$ in eq. (6b) as a function of the surface over volume ratio and of $P_{fast \atop nano}$, $P_{slow \atop nano}$, $P_{fast \atop macro}$, $P_{slow \atop macro}$ according to eq. (6a); note that $P_{fast \atop nano} \equiv P_{nano \atop fast}$ and so on, thus, synthetically, the condition of symmetry $P_{ab} \equiv P_{ba}$ holds. Following the two different approaches we find the same result, as must be for self-consistency if and only if $\delta = P_{slow \atop nano} / P_{slow \atop macro} = P_{fast \atop nano} / P_{fast \atop macro}$, i.e., synthetically, $P_{ab}/P_{ac} = P_{db}/P_{dc} \equiv P_b/P_c$. Accordingly:

$$P(S/V, t) = P_{\substack{nano \\ fast}} \left( \frac{(P_{nano}/P_{macro})^\gamma - 1}{\ell S/V + 1} + 1 \right)^{-1/\gamma} \left( \frac{(P_{fast}/P_{slow})^\nu - 1}{\tau/t + 1} + 1 \right)^{-1/\nu} \quad (6c)$$

Eq. (6c) is the spatial-temporal universal scaling law that we propose; note that accordingly to the self-consistent condition we require in addition $P_{ab} = P_{ac}P_b/P_c$; thus, all the limit cases of eqs. (6a) and (6b) are recovered. The generalization of eq. (6c) to include new parameters in addition to $S/V$ and $t$ (e.g., a velocity) is evident. One example of application is on the friction coefficient, for which a spatial-temporal scaling is observed ("fast" and "slow" here would define the "static" and "dynamic" friction coefficients); note that our approach quantifies also the influence of the indenter shape (and not only of its size), e.g., on the friction coefficient.

Concluding, the universality of the derived spatial-temporal scaling could have large applicability in different fields, even for complex, chaotic and fractal systems, as present in engineering, physics, biology, medicine, economy, to cite a few; and obviously in nanoindentation.

Acknowledgements
The author would like to thank Profs. A. Carpinteri and P. P. Delsanto for discussion and Renee Eaton for the English grammar supervision.

FIGURES

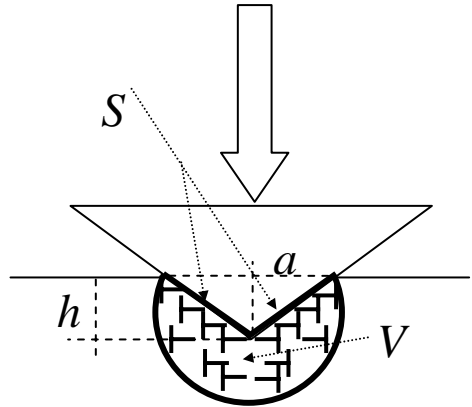

Figure 1: Geometrical necessary dislocations during indentation: $h$ is the indentation depth, $a$ is the radius of the projected contact area, $S$ is the contact surface, and $V$ is the dissipation domain (proportional to $a^3$). Note that the indented surface at the nanoscale appears in discrete steps due to the formation of dislocation loops, i.e., of quantized plasticity. In our model the scaling law is predicted to be a function only of $S/V$.

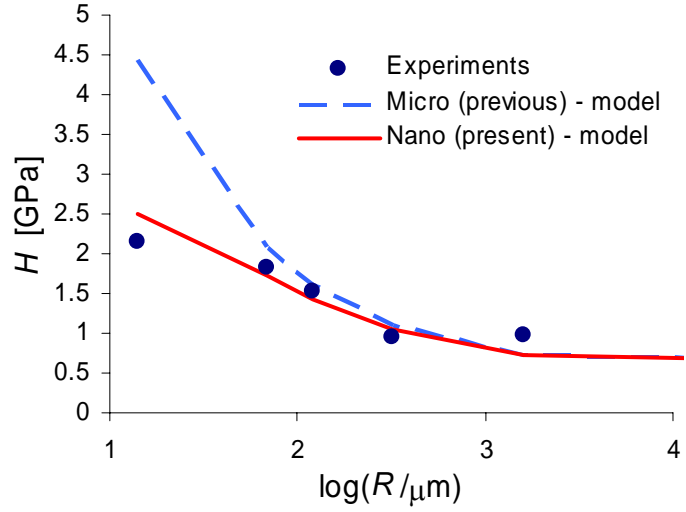

(a)

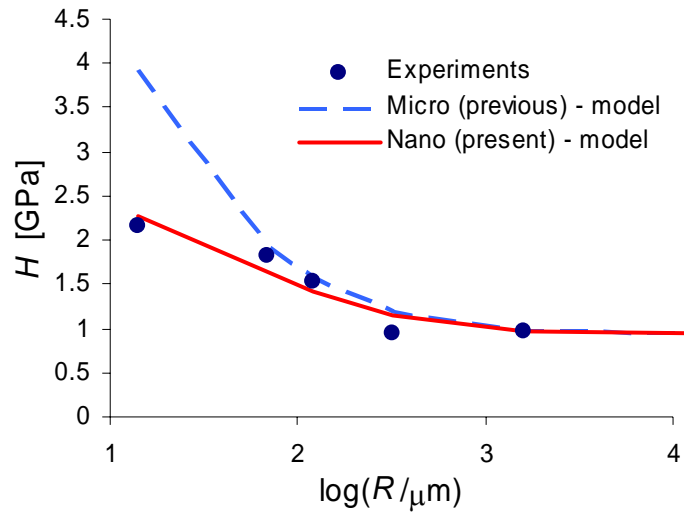

(b)

Figure 2: Comparison between "micro-models" (Nix and Gao, 1998 and Swadener et al., 2002) and present "nano-model" fitted to experiments on spherical indentation by Swadener et al. (2002). They considered two different sets of parameters: (a) $H_{macro} \approx 0.9\text{GPa}$, $R^* \approx 250\mu\text{m}$ (dotted line), that introduced in the nano-model (solid line) with $H_{nano} \approx 3\text{GPa}$ result in a closer agreement with the experiments (points); (b) same, but for $H_{macro} \approx 0.6\text{GPa}$, $R^* \approx 750\mu\text{m}$ (dotted line), or in addition $H_{nano} \approx 5\text{GPa}$ (solid line). Note the difference in the predictions of the models for the smaller size-scale.